\documentclass[prc,aps,preprintnumbers,nofootinbib,amsmath,amssymb]{revtex4}
\usepackage{epsfig}

\newcommand{\be}{\begin{equation}}
\newcommand{\ee}{\end{equation}}
\newcommand{\bea}{\begin{eqnarray}}
\newcommand{\eea}{\end{eqnarray}}

\begin{document}
\title{ Collective Flow signals the Quark Gluon Plasma
\footnote{Supported by DFG, GSI, BMBF, EU, RIKEN and DOE} }

\author{H.~St\"ocker}
\vspace*{5mm}
 \affiliation{Institut f\"{u}r Theoretische Physik,
 Johann Wolfgang Goethe -- Universit\"{a}t,
 Robert Mayer Str. 8-10,
 60054 Frankfurt am Main, Germany}
 \affiliation{
 Frankfurt Institute for Advanced Studies (FIAS),
 Robert Mayer Str. 8-10,
 60054 Frankfurt am Main, Germany}


\begin{abstract}
A critical discussion of the present status of the CERN experiments on
charm dynamics and hadron collective flow is given. We emphasize the
importance of the flow excitation function from 1 to 50 A$\cdot$GeV:
here the hydrodynamic model has predicted the collapse of the
$v_1$-flow  and of the $v_2$-flow at $\sim 10$ A$\cdot$GeV; at 40
A$\cdot$GeV it has been recently observed by the NA49 collaboration.
Since hadronic rescattering models predict much larger flow than
observed at this energy we interpret this observation as potential
evidence for a first order phase transition at high baryon density
$\rho_B$.  A detailed discussion of the collective flow as a barometer
for the equation of state (EoS) of hot dense matter at RHIC follows.
Here, hadronic rescattering models can explain $< 30 \%$ of the
observed elliptic flow, $v_2$, for $p_T > 2$ GeV/c.  This is
interpreted as evidence for the production of superdense matter at RHIC
with initial pressure far above hadronic pressure, $p > 1$ GeV/fm$^3$.
We suggest that the fluctuations in the flow, $v_1$ and $v_2$, should
be measured in future since ideal hydrodynamics predicts that they are
larger than 50 \%  due to initial state fluctuations.  Furthermore, the
QGP coefficient of viscosity may be determined experimentally from the
fluctuations observed.  The connection of $v_2$ to jet suppression is
examined. It is proven experimentally that the collective flow is not
faked by minijet fragmentation. Additionally, detailed transport
studies show that the away-side jet suppression can only partially ($<$
50\%) be due to hadronic rescattering. We, finally, propose upgrades
and second generation experiments at RHIC which inspect the first order
phase transition in the fragmentation region, i.e. at $\mu_B \approx
400$ MeV ($y \approx 4-5$), where the collapse of the proton flow
should be seen in analogy to the 40 A$\cdot$GeV data.  The study of
Jet-Wake-riding potentials and Bow shocks -- caused by jets in the QGP
formed at RHIC -- can give further information on the equation of state
(EoS) and transport coefficients of the Quark Gluon Plasma (QGP).
\end{abstract}

\maketitle


\section{Old and new observables for the QGP phase
transition}

\begin{figure}[t]
\begin{center}
\centerline{\epsfig{file=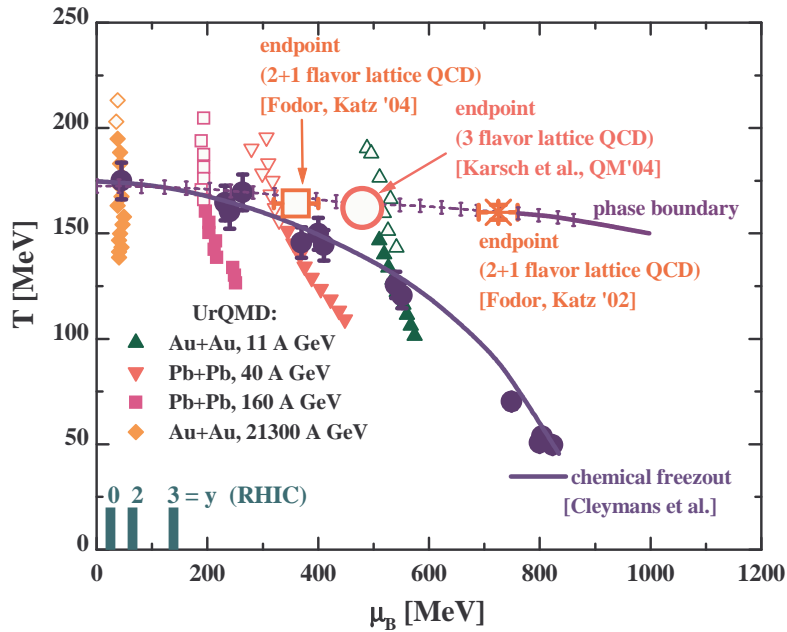,scale=0.55}}
\caption{
The new phase diagram with the critical end point at $\mu_B \approx 400
\mbox{ MeV}, T \approx 160 \mbox{ MeV} $ as predicted by Lattice QCD.
In addition, the time evolution in the $T-\mu$-plane of a central cell
in UrQMD calculations \cite{Bravina} is depicted for
different bombarding energies.  Note, that the calculations indicate
that bombarding energies $E_{LAB} \stackrel{<}{\sim} 40$ A$\cdot$GeV are
needed to probe a first order phase transition. At RHIC (see insert at
the $\mu_B$ scale) this point is accessible in the fragmentation region
only (taken from \protect{\cite{Bratkov04}}).}
\label{phasedia}
\end{center}
\end{figure}

Lattice QCD results \cite{Fodor04,Karsch04} (cf. Fig.  \ref{phasedia})
show a crossing, but no first order phase transition to the QGP for
vanishing or small chemical potentials $\mu_B$, i.e. at the conditions
accessible at central rapidities at RHIC full energies.  A first order
phase transition does occur according to the QCD lattice calculations
\cite{Fodor04,Karsch04} only at high baryochemical potentials or
densities, i.e.  at SIS-300 and lower SPS energies and in the
fragmentation region of RHIC, $y \approx 4-5$
\cite{Anishetty80,Date85}.  The critical baryochemical potential is
predicted \cite{Fodor04,Karsch04} to be $\mu_B^c \approx 400 \pm 50
\mbox{ MeV} $ and the critical temperature $T_c \approx 150 - 160$ MeV.
We do expect a phase transition also at finite strangeness.
Predictions for the phase diagram of strongly interacting matter for
realistic non-vanishing net strangeness are urgently needed to obtain a
comprehensive picture of the QCD phase structure.  Multi-Strangeness
degrees of freedom are very promising probes for the properties of the
dense and hot matter \cite{Koch86}. The strangeness distillation
process \cite{greic87,greic88} predicts dynamical de-admixture of $s$
and $\bar{s}$ quarks, which yields unique signatures for QGP creation:
high multistrange hyperon-/-matter production, strangelet formation and
unusual antibaryon to baryon ratios $ect$.

\subsection{Thermodynamics in the $T-\mu_B$ plane}

A comparison of the thermodynamic parameters $T$ and $\mu_B$ extracted
from the UrQMD-transport model in the central overlap regime of Au+Au
collisions \cite{Bratkov04} with the QCD predictions is shown in Fig 1,
where the full dots with errorbars denote the 'experimental' chemical
freeze-out parameters -- determined from fits to the experimental
yields -- taken from Ref.  \cite{Cleymans}. The triangular and
quadratic symbols (time-ordered in vertical sequence) stand for
temperatures $T$ and chemical potentials $\mu_B$ extracted from UrQMD
transport calculations in central Au+Au (Pb+Pb) collisions at RHIC
(21.3 A$\cdot$TeV), 160, 40 and 11 A$\cdot$GeV \cite{Bravina} as a
function of the reaction time (separated by 1 fm/c steps from top to
bottom).  The open symbols denote nonequilibrium configurations and
correspond to $T$ parameters extracted from the transverse momentum
distributions, whereas the full symbols denote configurations in
approximate pressure equilibrium in longitudinal and transverse
direction.

During the nonequilibrium phase (open symbols) the transport
calculations show much higher temperatures (or energy densities) than
the 'experimental' chemical freeze-out configurations at all bombarding
energies ($\geq$ 11 A$\cdot$GeV).  These numbers are also higher than
the critical point (circle) of (2+1) flavor - Lattice QCD calculations
by the Bielefeld-Swansea-collaboration  \cite{Karsch04} (large open
circle) and by the Wuppertal-Budapest-collaboration \cite{Fodor04} (the
star shows earlier results from \cite{Fodor04}).  The energy density at
$\mu_c, T_c$ is in the order of  $\approx$ 1 GeV/fm$^3$ (or slightly
below).  At RHIC energies a cross-over is expected at midrapidity, when
stepping down in temperature during the expansion phase of the 'hot
fireball'. The baryon chemical potential $\mu_B$ for different rapidity
intervals at RHIC energies has been obtained from a statistical model
analysis by the BRAHMS Collaboration based on measured antihadron to
hadron yield ratios \cite{BRAHMS_PRL03}.  For midrapidity one finds
$\mu_B\simeq 0$, whereas for forward rapidities $\mu_B$ increases up to
$\mu_B\simeq 130$~MeV at $y=3$.  Thus, only extended forward rapidity
measurement ($y \approx 4-5)$ will allow to probe large $\mu_B$ at
RHIC.  The detectors at RHIC at present offer only a limited chemical
potential range. To reach the first order phase transition region at
midrapidity, the International Facility at GSI seems to be the right
place to go. This situation changes at lower SPS (and top AGS) as well
as at the future GSI SIS-300 energies: sufficiently large chemical
potentials $\mu_B$ should allow for a first order phase transition
\cite{Shuryak} (to the right of the critical point in the ($T, \mu_B$)
plane).  The transport calculations show high temperatures (high energy
densities) in the very early phase of the collisions, only. Here,
hadronic interactions are weak due to formation time effects and yield
little pressure.  Diquark, quark and gluon interactions should cure
this problem.

\subsection{{ Open charm and charmonia at SPS and RHIC}}

Open charm and charmonium production at SPS and RHIC energies has been
calculated within the HSD and UrQMD transport approaches
\cite{Cassing03} using parametrizations for the elementary production
channels including the charmed hadrons $D, \bar{D}, D^*, \bar{D}^*,
D_s, \bar{D}_s, D_s^*, \bar{D}_s^*,$ $J/\Psi, \Psi(2S), \chi_{2c}$ from
$NN$ and $\pi N$ collisions. The latter parametrizations are fitted to
PYTHIA calculations \cite{PYTHIA} above $\sqrt{s}$ = 10 GeV and
extrapolated to the individual thresholds, while the absolute strength
of the cross sections is fixed by the experimental data as described in
Ref.  \cite{Cass01}; for previous works see
\cite{Cass01a,Cass97,Spieles} Backward channels 'charm + anticharm
meson $\rightarrow$ charmonia + meson' are treated via detailed balance
in a more schematic interaction model with a single parameter or matrix
element $|M|^2$ that is fixed by the $J/\Psi$ suppression data from the
NA50 collaboration at SPS energies (cf. Ref. \cite{Brat03}).

\begin{figure}[t]
\centerline{\epsfig{file=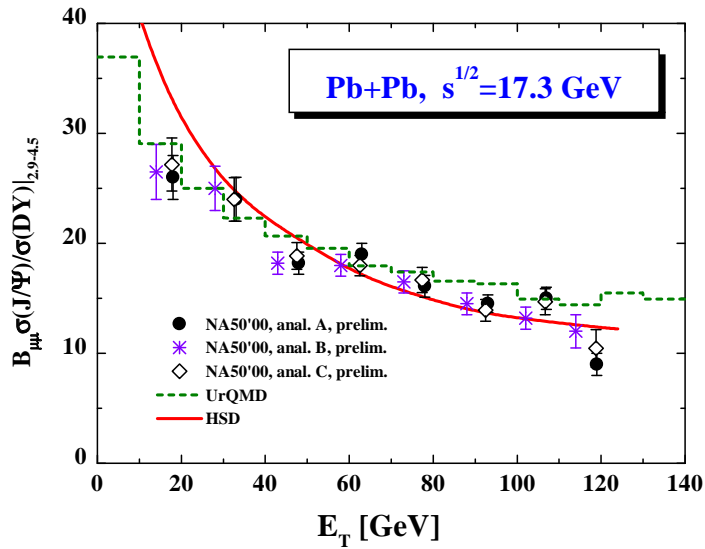,scale=0.6}}
\caption{The $J/\Psi$ suppression as a function of the transverse
energy $E_T$ in $Pb~+~Pb$ collisions at 160 A$\cdot$GeV. The solid line shows
the HSD result within the comover absorption scenario
\protect\cite{Brat03}. The different symbols stand for the NA50 data
\protect\cite{NA50_QM02} from the year 2000 (ana\-lysis A,B,C) while
the dashed histogram is the UrQMD result \protect\cite{Spieles}.}
\label{Fig2} \end{figure}

\begin{figure}[b]
\centerline{\epsfig{file=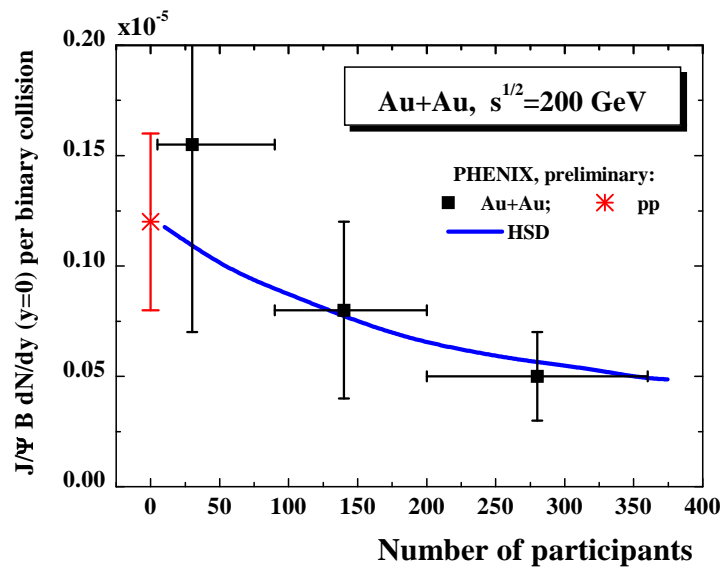,scale=0.6}}
\caption{The calculated
$J/\Psi$ multiplicity per binary collision -- multiplied by the
branching to dileptons --  as a function of the number of
participating nucleons, $N_{part}$, in comparison to the preliminary
data from the PHENIX collaboration \protect\cite{PHENIX2} for
$Au+Au$ and $pp$ reactions (taken from Ref. \protect\cite{Brat03}).}
\label{bild14n}
\end{figure}

We recall that charmonium suppression had been proposed as ''the
clearest'' QGP-signature and the community has been riding on this
folklore for almost two decades.  Hence, these detailed comparisons
come as a shock:  The $Pb+Pb$ results at 160 A$\cdot$GeV, both from
UrQMD and HSD transport calculations are well in line with the data of
the NA50 Collaboration in Fig. \ref{Fig2}, where the $J/\Psi$
suppression is shown as a function of the transverse energy $E_T$.  The
solid line stands for the HSD result within the  comover absorption
scenario while the various data points reflect the  NA50 data from the
year 2000 (analysis A,B,C).  The data have moved so that they agree now
with the HSD and UrQMD calculations \cite{Spieles} (dashed
histogram in Fig.  \ref{Fig2}). We mention that there might be
alternative explanations for $J/\Psi$ suppression, as discussed in
Refs. \cite{Satz99,Rappnew} and/or further (dissociation) mechanisms
not considered here. However, for the purposes of the present review it
is sufficient to point out that

a) no sign of unusual physics can be related to the $J/\Psi$ data and

b) the models employed here (cf. Figs. 6 and 7 in \cite{Brat03}) use
upper limits for the dissociation cross sections and do not lead to a
sizeable re-creation of charmonia by $D+\bar{D}$ channels at SPS
energies.

At RHIC  central $Au+Au$ collisions at $\sqrt{s}$ = 200 GeV will,
however, produce multiplicities of open charm pairs about 2 orders of
magnitude larger than at 160 A$\cdot$GeV, such that a much higher
$J/\Psi$ reformation rate ($\sim N_{c\bar{c}}^2$) is expected (cf. Ref.
\cite{Rappnew}).  At RHIC top energy, $\sqrt{s}$ = 200 GeV,  the
$J/\Psi$ comover dissociation may no longer be important, since the
charmonia dissociated in this channel are approximately recreated in
the backward channels. Hence, the $J/\Psi$ dissociation at RHIC should
be less pronounced  than at SPS energies.

The preliminary data of the PHENIX Collaboration \cite{PHENIX2} allow
for a first glance at the situation encountered in $Au+Au$ collisions
at $\sqrt{s}$ = 200 GeV.  Fig. \ref{bild14n} shows the $J/\Psi$
multiplicity per binary collision as a function of the number of
participating nucleons $N_{part}$ in comparison to the data at
midrapidity. The statistics is quite limited; thus no final conclusion
can presently be drawn. However, the data neither suggest a dramatic
enhancement of $J/\Psi$ production nor a complete 'melting' of the
charmonia in the QGP phase.

\subsection{Historical Interlude}

Hydrodynamic flow and shock formation has been proposed early
\cite{Hofmann74,Hofmann76} as the key mechanism for the creation of
hot and dense matter during relativistic heavy-ion collisions.  The
full three-dimensional hydrodynamical flow problem is much more
complicated than the one-dimensional Landau model \cite{Landau}:  the
3-dimensional compression and expansion dynamics yields complex triple
differential cross-sections, which provide quite accurate spectroscopic
handles on the equation of state.  The bounce-off, $v_1(p_T)$, the
squeeze-out, $v_2(p_T)$, and the antiflow
\cite{Stocker79,Stocker80,Stocker81,Stocker82,Stocker86} (third flow
component \cite{Csernai99,Csernai04}) serve as differential barometers
for the properties of compressed, dense matter from SIS to RHIC.  The
most employed flow observables are:
\bea v_1 &=& \left< \frac{p_x}{p_T} \right> \\ v_2 &=& \left<
\frac{p_x^2 - p_y^2}{p_x^2 + p_y^2} \right> \, .
\eea
Here, $p_x$ denotes the momentum in
$x$-direction, i.e. the transversal momentum within the reaction plane
and $p_y$ the transversal momentum out of the reaction plane. The total
transverse momentum is given as $p_T = \sqrt{p_x^2 + p_y^2}$; the
$z$-axis is in the beam direction.  Thus, flow $v_1$ measures the
''bounce-off'', i.e. the strength of the directed flow in the reaction
plane, and $v_2$ gives the strength of the second moment of the
azimuthal particle emission distribution, the so-called
''squeeze-out''.
\cite{Hofmann74,Hofmann76,Stocker79,Stocker80,Stocker81,Stocker82,Stocker86}.
In particular,  it has been shown
\cite{Hofmann76,Stocker79,Stocker80,Stocker81,Stocker82,Stocker86} that
the disappearence or ''collapse'' of flow is a direct result of a first
order phase transition.

Several hydrodynamic models have been used in the past, starting with
the one-fluid ideal hydrodynamic approach.  It is well known that the
latter model predicts far too large flow effects. To obtain a better
description of the dynamics, viscous fluid models have been developed
\cite{Schmidt93,Muronga01,Muronga03}.  In parallel, so-called
three-fluid models, which distinguish between projectile, target and
the fireball fluids, have been considered \cite{Brachmann97}.  Here
viscosity effects appear only between the different fluids, but not
inside the individual fluids. The aim is to have at our disposal a
reliable, three-dimensional, relativistic three-fluid model including
viscosity \cite{Muronga01,Muronga03}.

Flow can be described very elegantly in hydrodynamics. However, also
consider microscopic multicomponent (pre-) hadron transport theory,
e.g.  models like qMD \cite{Hofmann99}, IQMD \cite{Hartnack89}, UrQMD
\cite{Bass98} or HSD \cite{Cassing99}, as control models for viscous
hydro and as background models to subtract interesting non-hadronic
effects from data.  If Hydro with and without quark matter EoS,
hadronic transport models without quark matter -- but with strings --
are compared to data, can we learn whether quark matter has been
formed?  What degree of equilibration has been reached? What does the
equation of state look like?  How are the particle properties, self
energies, cross sections changed?

To estimate systematic model uncertainties, the results of the
different microscopic transport models also have to be carefully
compared. The two robust hadron/string based models, HSD and UrQMD, are
considered in the following.

\subsection{Review of AGS and SPS results}

Microscopic (pre-)hadronic transport models describe the formation and
distributions of many hadronic particles at AGS and SPS rather well
\cite{Weber02}.  Furthermore, the nuclear equation of state has been
extracted by comparing to flow data which are described reasonably well
up to AGS energies
\cite{Andronic03,Andronic01,Soff99,Csernai99,Sahu1,Sahu2}.  Ideal hydro
calculations, on the other hand, predict far too much flow at these
energies \cite{Schmidt93}.  Thus, viscosity effects have to be taken
into account in hydrodynamics.

\begin{figure}[t]
\begin{center}
\begin{minipage}[l]{6.5 cm}
\hspace{-1cm}
\epsfig{file=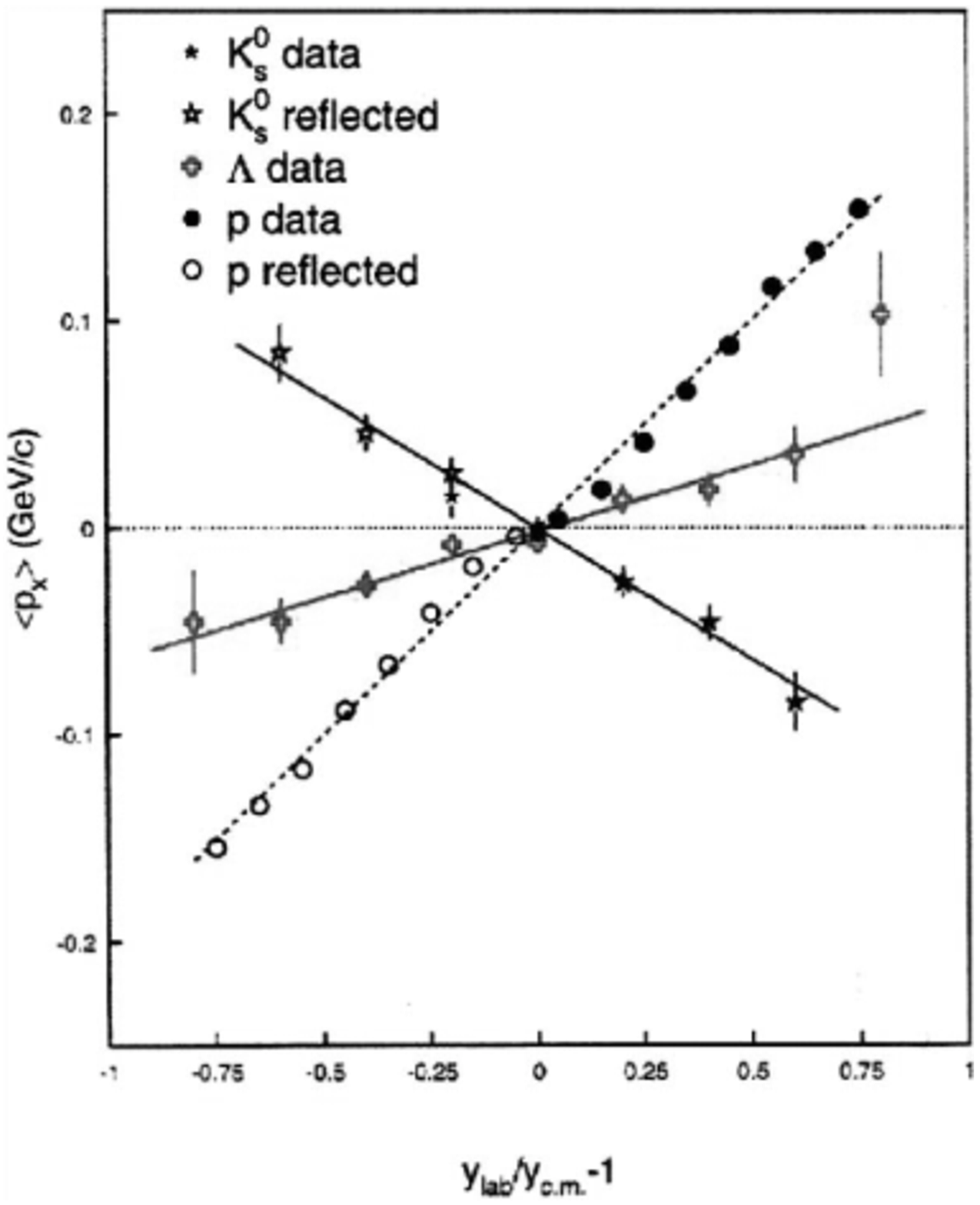,scale=0.47}
\end{minipage}
\begin{minipage}[l]{10 cm}
\vspace{-0.8cm}
\epsfig{file=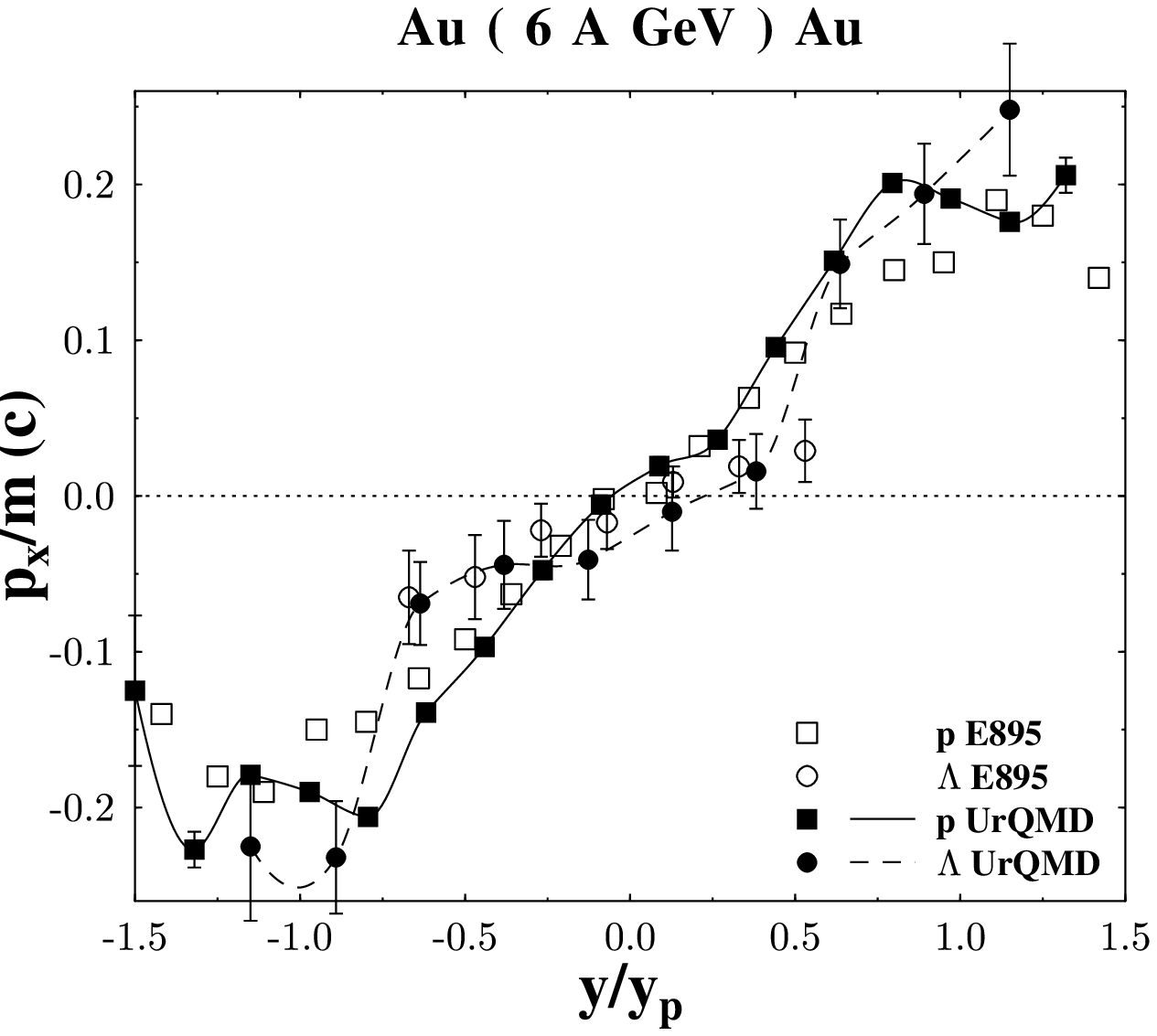,scale=0.75}
\end{minipage}
\caption{
Left: Sideward flow $p_x=v_1 \cdot p_T$ of K, $\Lambda$ and p's
at 6 A$\cdot$GeV  as
measured by E895 in semi-central collisions at the AGS.
Right: The same directed flow data for $p$ and $\Lambda$ compared to
UrQMD calculations for $b < 7$ fm
\protect{\cite{Soff99} }. }
\label{flow_ags_soff}
\end{center}
\end{figure}

\begin{figure}[t]
\centerline{\epsfig{file=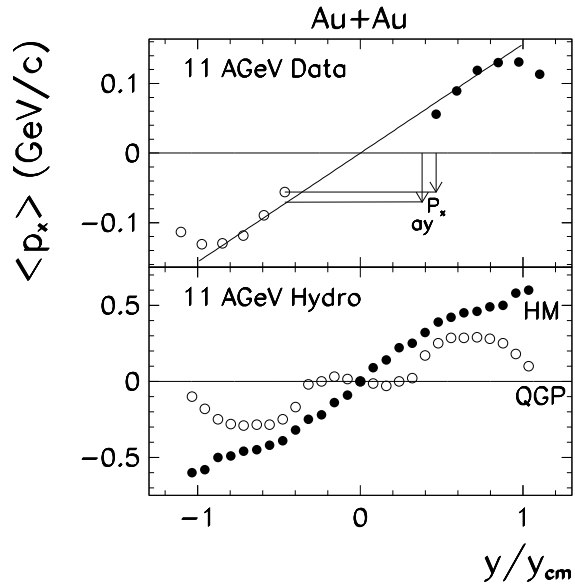,scale=0.45}}
\caption{
Directed flow from ideal hydrodynamics with a QGP phase (open symbols)
and from the Quark Gluon String Model without QGP phase (full symbols)
\protect{\cite{Csernai99} }. \label{flow_csernai}}
\end{figure}

\noindent
In particular, ideal hydro calculations are factors of two higher than
the measured sideward flow at SIS \cite{Schmidt93} and AGS, while the
directed flow $p_x/m$ measurement of the E895 collaboration shows that
the $p$ and $\Lambda$ data are reproduced reasonably well \cite{Soff99}
(Fig. \ref{flow_ags_soff}) in UrQMD, i.e. in a hadronic transport
theory with reasonable cross-sections, i.e.  realistic mean-free-path
of the constituents.

\begin{figure}[t]
\epsfig{file=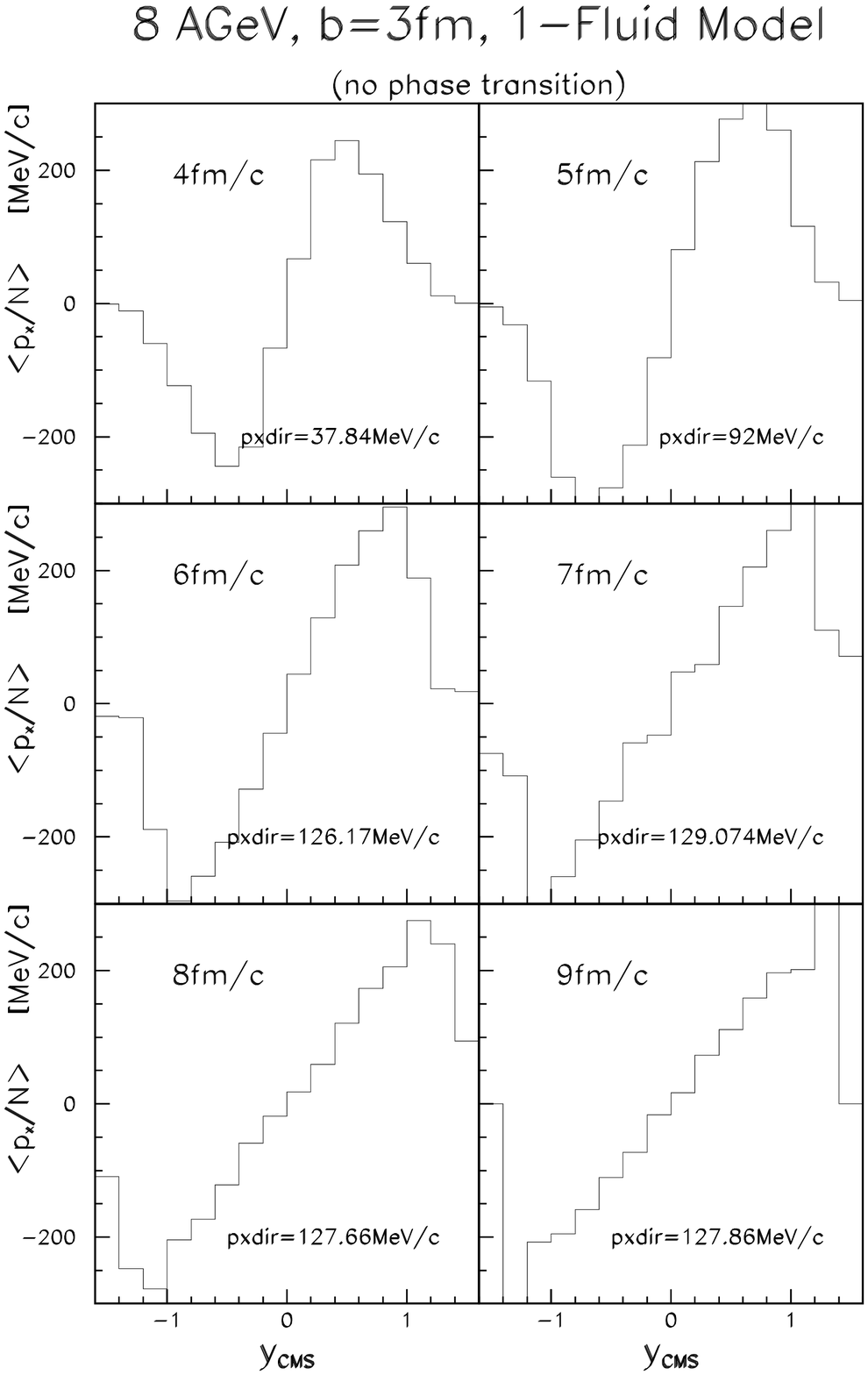,width=7cm}\hspace*{10mm}
\epsfig{file=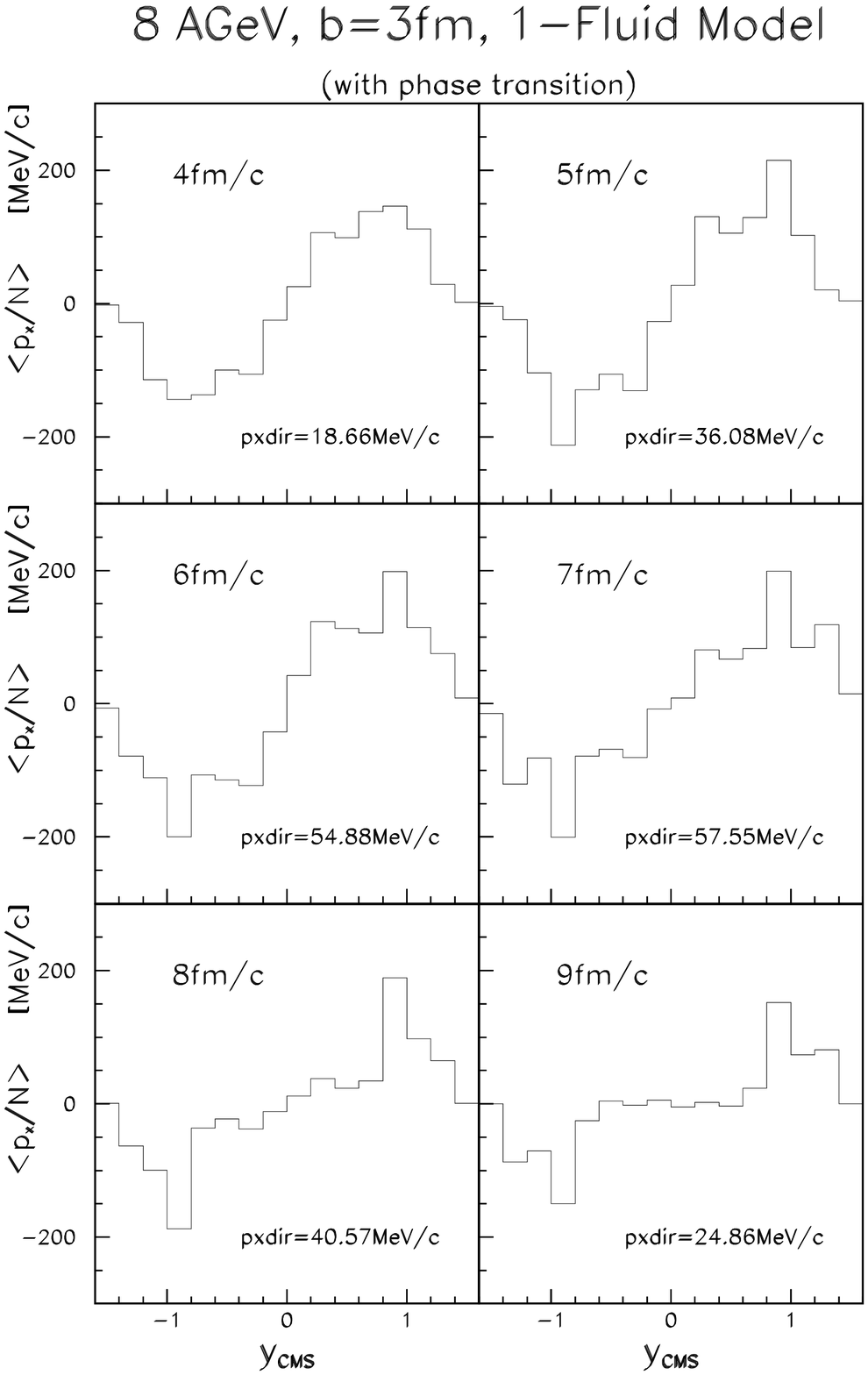,width=7cm}
\caption{
Time evolution of directed flow $p_x/N$ as a function of rapidity for
Au+Au collisions at 8 A$\cdot$GeV in the one-fluid model.
Left: Hadronic EoS without phase
transition. Right: An EoS including a first order phase transition to
the QGP \protect{\cite{Brach00}}. \label{flow_brach1}}
\end{figure}

Only ideal hydro calculations predict, however, the appearance of a
so-called ''third flow component'' \cite{Csernai99} or ''antiflow''
\cite{Brach00} in central collisions (cf. Fig. \ref{flow_csernai}). We
stress that this only holds if the matter undergoes a first order phase
transition to the QGP. The signal is that around midrapidity the
directed flow, $p_x (y)$, of protons develops a negative slope! In
contrast, a hadronic EoS without QGP phase transition does not yield
such an exotic ''antiflow'' (negative slope) wiggle in the proton flow
$v_1(y)$.  The ideal hydrodynamic time evolution of the directed flow,
$p_x/N$, for the purely hadronic EoS (Fig.  \ref{flow_brach1} l.h.s.)
does show a clean linear increase of $p_x(y)$, just as the microscopic
transport theory (Fig. \ref{flow_ags_soff} r.h.s.) and as the data
(Fig.  \ref{flow_ags_soff} l.h.s.). For an EoS including a first order
phase transition to the QGP (Fig.  \ref{flow_brach1} r.h.s.) it can be
seen, however,  that the proton flow $v_1 \sim p_x/p_T$ collapses; the
collapse occurs around midrapidity.

\noindent
This observation is explained by an antiflow component of protons,
which develops when the expansion from the plasma sets in
\cite{Brach99}.

\begin{figure}[!]
\phantom{a}\vspace*{-15mm}
\centerline{\epsfig{file=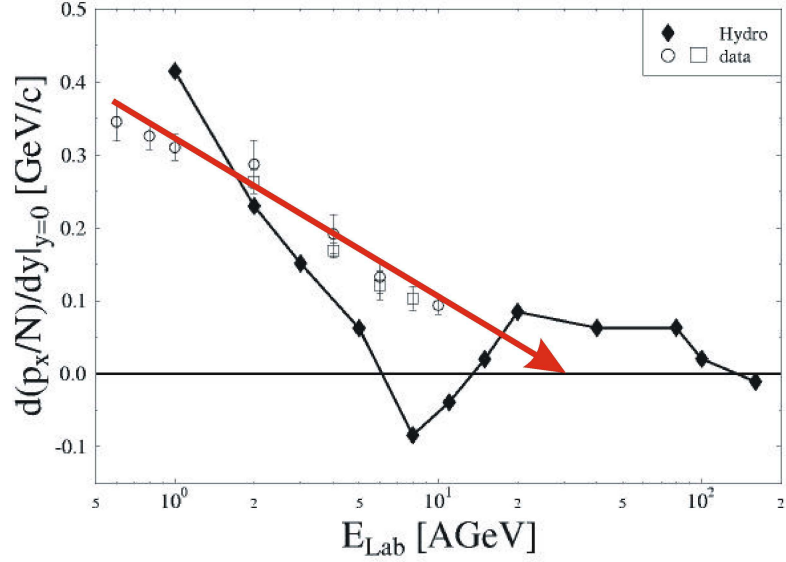,scale=0.5}}
\caption{Measured SIS and AGS proton $dp_x/dy$-slope
data compared to a one-fluid hydro calculation.
A linear extrapolation of the AGS data indicates a collapse of flow at
$E_{Lab} \approx 30$ A$\cdot$GeV, i.e. for the lowest SPS- and the upper
FAIR- energies at GSI \protect{\cite{Paech00}.}
\label{flow_extra}}
\end{figure}

The ideal hydrodynamic directed proton flow $p_x$  (Fig.
\ref{flow_extra}) shows even negative values between 8 and 20
A$\cdot$GeV. An increase back to positive flow is predicted with
increasing energy, when the compressed QGP phase is probed.  But, where
is the predicted minimum of the proton flow in the data?  The hydro
calculations suggest this ''softest point collapse'' is at $E_{Lab}
\approx 8$ A$\cdot$GeV.  This has not been verified by the AGS data!
However, a linear extrapolation of the AGS data indicates a collapse of
the directed proton flow at $E_{Lab} \approx 30$ A$\cdot$GeV (Fig.
\ref{flow_extra}).

\begin{figure}[h]
\begin{center}
\begin{minipage}[l]{12 cm}
\epsfig{file=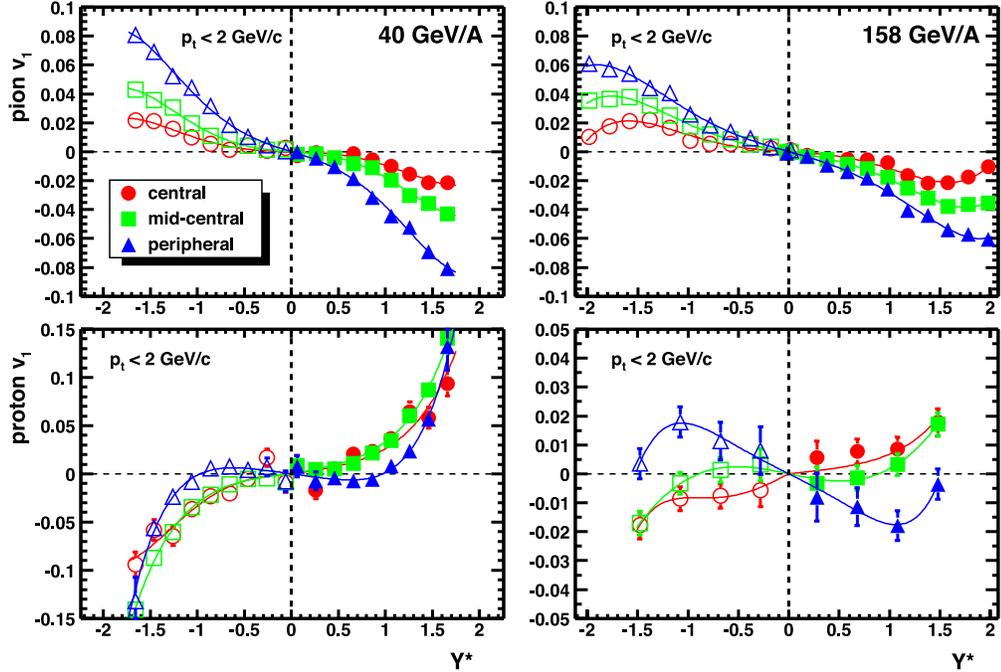,scale=0.75}
\end{minipage}
\caption{ $v_1$ at SPS, 40 A$\cdot$GeV and 158 A$\cdot$GeV
\protect{\cite{NA49_v2pr40} }. The proton antiflow is observed in the
NA49-experiment even at near central collisions, which is in contrast
to the UrQMD-model involving no phase transition
(Fig. \protect\ref{v1_sps40}). }
\label{sps_v1_data}
\end{center}
\end{figure}

\begin{figure}[h]
\begin{center}
\epsfig{file=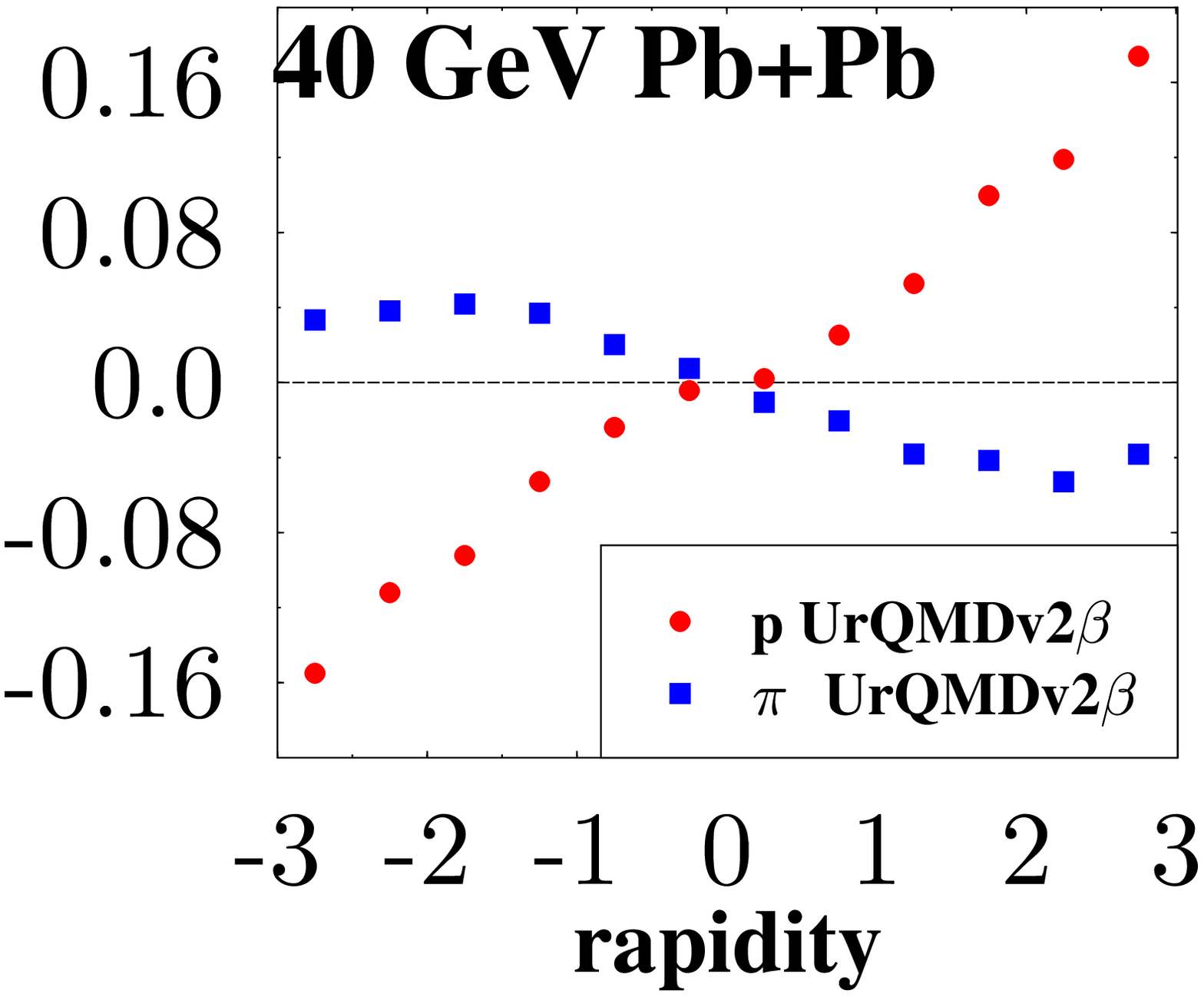,scale=0.50}
\caption{Proton and pion flow
$v_1=p_x/p_T$  at 40 A$\cdot$GeV as obtained within the UrQMD model.
No proton antiflow is generated in this hadronic transport theory without
phase transition.}
\label{v1_sps40}
\end{center}
\end{figure}

Recently, substantial support for this prediction has been obtained by
the low energy 40 A$\cdot$GeV SPS data of the NA49 collaboration
\cite{NA49_v2pr40} (cf. Fig. \ref{sps_v1_data}).  These data clearly
show the first proton ''antiflow'' around mid-rapidity, in contrast to
the AGS data as well as to the UrQMD calculations involving no phase
transition (Fig. \ref{v1_sps40}).
Thus, at bombarding energies of 30-40 A$\cdot$GeV, a first order phase
transition to the baryon rich QGP most likely is observed; the first
order phase transition line is crossed (cf. Fig.  \ref{phasedia}). This
is the energy region where the new FAIR- facility at GSI will operate.
There are good prospects that the baryon flow collapses and other first
order QGP phase transition signals can be studied soon at the lowest
SPS energies as well as at the RHIC fragmentation region $y > 4-5$.
These experiments will enable a detailed study of the first order phase
transition at high $\mu_B$ and of the properties of the baryon rich
QGP.

\section{Proton elliptic flow collapse at 40 A$\cdot$GeV - evidence for a
first order phase transition at highest net baryon densities}

\begin{figure}[h]
\centerline{\epsfig{file=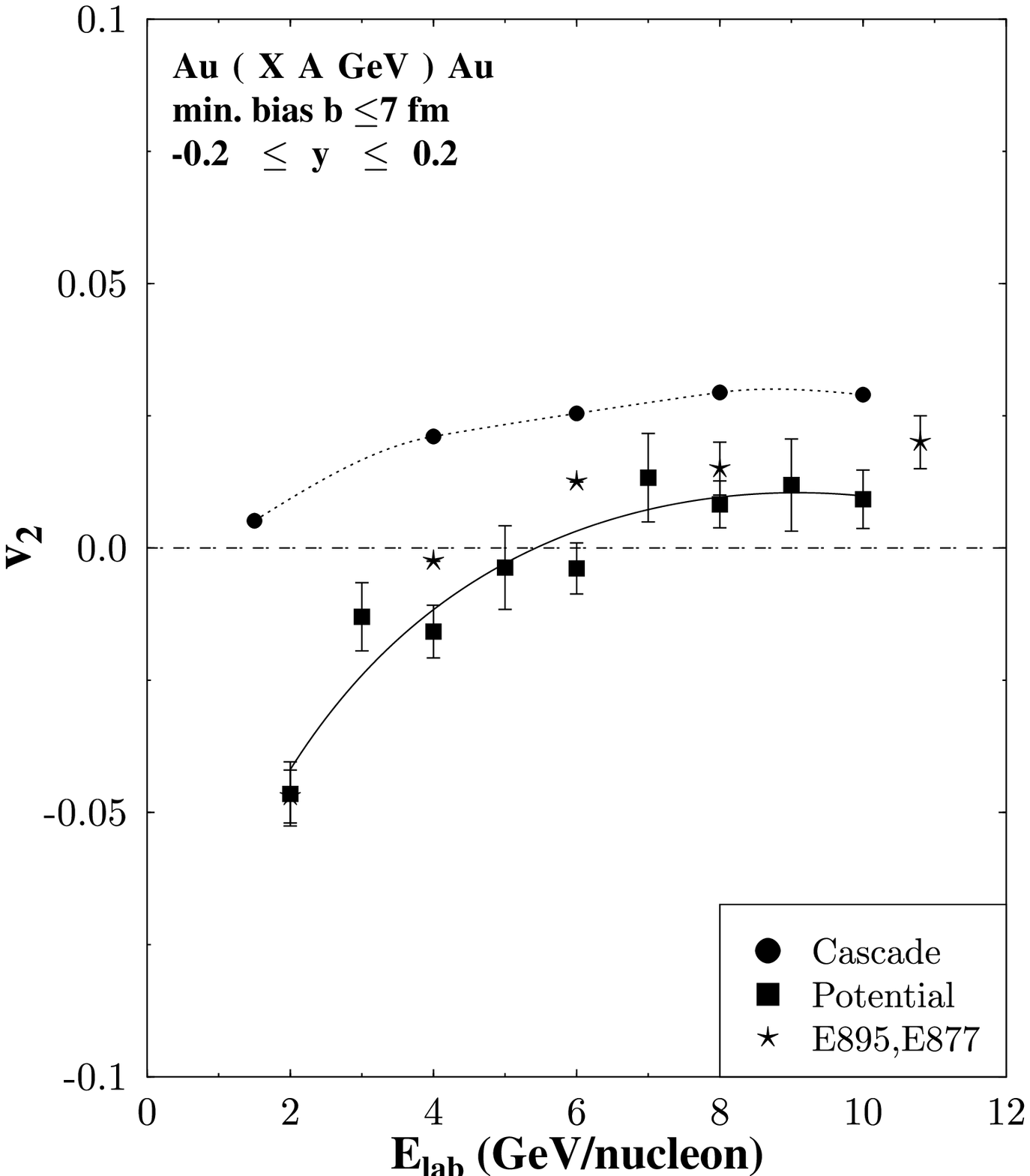,scale=0.4}}
\caption{
$v_2$ excitation function of protons at the AGS. The
E895-E877 data show the transition from squeeze-out to in-plane proton
elliptic flow at 4-5 A$\cdot$GeV; the UrQMD calculations show a strong
sensitivity to the EoS \protect\cite{Soff99}.}
\label{v2_excitation}
\end{figure}

\begin{figure}[h]
\centerline{\epsfig{file=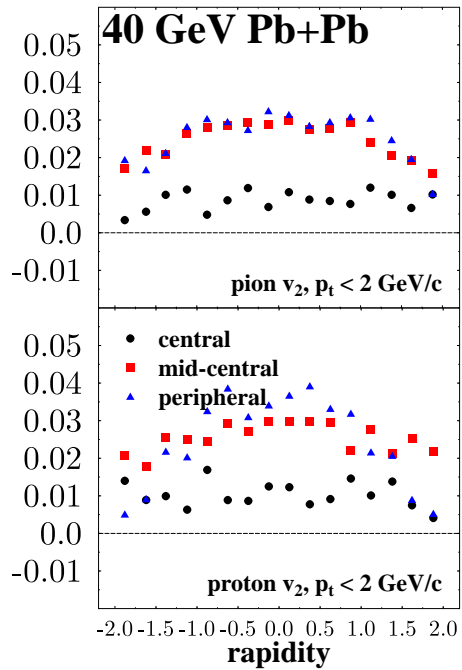,scale=0.4}}
\caption{Elliptic flow $v_2$ of protons (lower frame) and pions
(upper frame) versus rapidity for
Pb+Pb collisions at 40 A$\cdot$GeV from the UrQMD calculations
\protect{\cite{Soff99}}.}
\label{soff_v2pp40}
\end{figure}

\begin{figure}[h]
\centerline{\epsfig{file=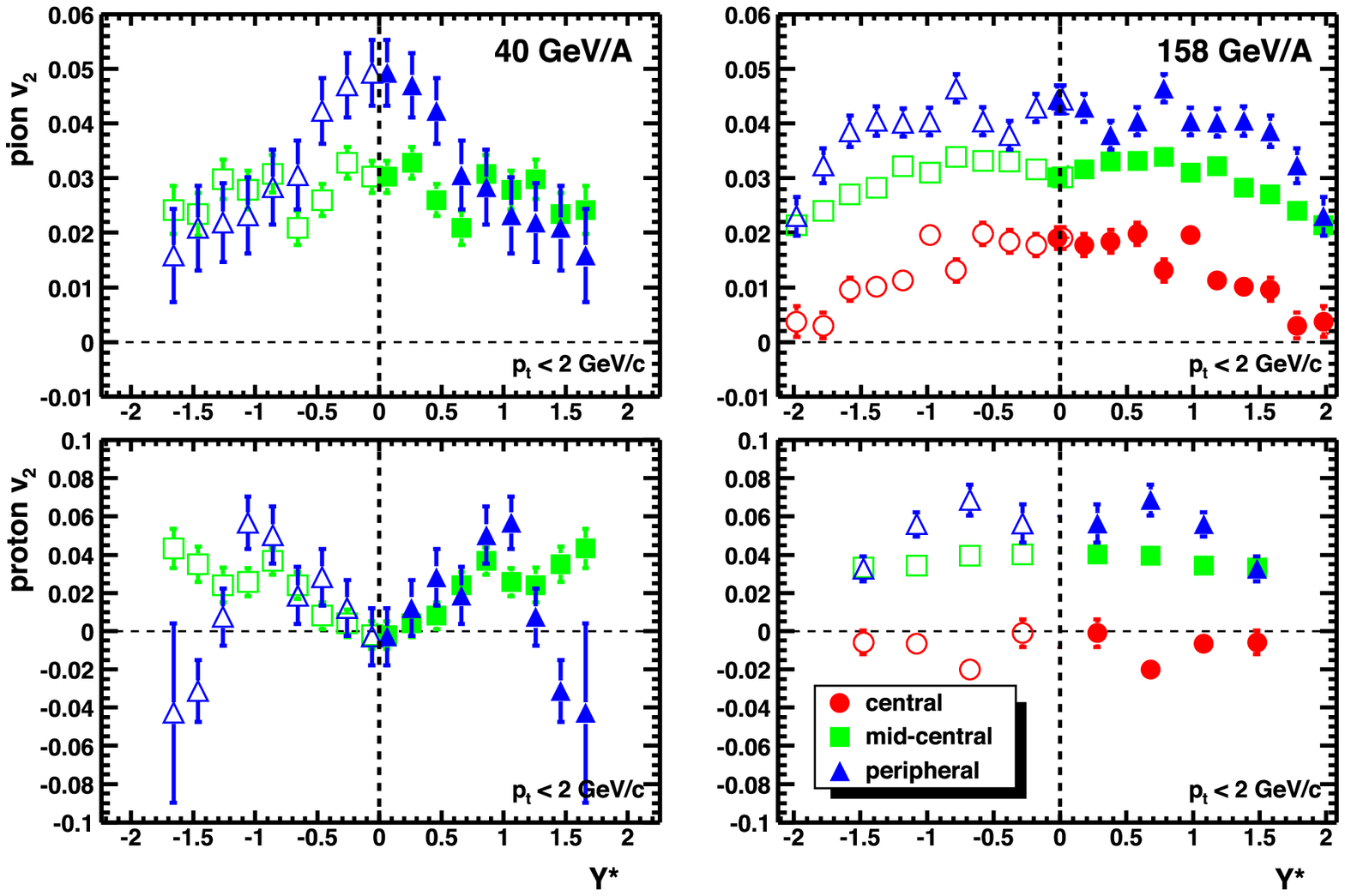,scale=0.6}}
\caption{Elliptic flow $v_2$ of protons  versus rapidity at 40
A$\cdot$GeV Pb+Pb collisions \protect\cite{NA49_v2pr40} as measured by NA49 for
three centrality bins:  central (dots), mid-central (squares) and
peripheral (triangles).  The solid lines are polynomial fits to the
data \protect\cite{NA49_v2pr40}.}
\label{Fig_v2pr40}
\end{figure}

At SIS energies microscopic transport models reproduce the data on the
excitation function of the proton elliptic flow $v_2$ quite well:  A
soft, momentum-dependent equation of state
\cite{Andronic00,Andronic99,Larionov} seems to account for the data.
The observed proton flow $v_2$ below $\sim$ 5 A$\cdot$GeV is smaller than
zero, which corresponds to the squeeze-out predicted by hydrodynamics
long ago
\cite{Hofmann74,Hofmann76,Stocker79,Stocker80,Stocker81,Stocker82,Stocker86}.
The AGS data (Fig. \ref{v2_excitation}) exhibit a transition from
squeeze-out to in-plane flow in the midrapidity region.  The change in
sign of the proton $v_2$ at 4-5 A$\cdot$GeV is in accord with transport
calculations (UrQMD calculations in Fig. \ref{v2_excitation} \cite{Soff99};
for HSD results see  \cite{Sahu1,Sahu2}).  At higher
energies, 10-160 A$\cdot$GeV, a smooth increase of the flow $v_2$ is predicted
from the hadronic transport simulations.  In fact, the 158 A$\cdot$GeV data of
the NA49-collaboration suggest that this smooth increase proceeds
between AGS and SPS as predicted.  Accordingly, UrQMD gives
considerable (~3\%) $v_2$ flow for midcentral and peripheral protons at
40 A$\cdot$GeV -- Fig. \ref{soff_v2pp40} \cite{Soff99}.

This is in strong contrast to recent NA49 data at 40 A$\cdot$GeV (cf.
Fig. \ref{Fig_v2pr40}): A sudden collapse of the proton flow is
observed for midcentral as well as for peripheral protons.  This
collapse of $v_2$ for protons around midrapidity at 40 A$\cdot$GeV is
very pronounced while it is not observed at 158 A$\cdot$GeV. The UrQMD
calculations, without a phase transition, show a robust, but wrong ~3\%
flow of protons - in strong contrast to the data.

\begin{figure}[t!]
\centerline{\epsfig{file=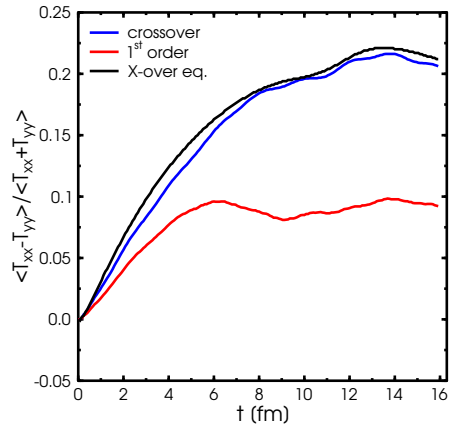,scale=0.35}}
\caption{Time evolution of the momentum anisotropy $v_2$ for a cross
over EoS. Both, in- and off-equilibrium calculations (upper lines) show
no drop of the flow  $v_2$.  The calculation with a first order phase
transition off-equilibrium (lower line) shows - on the other hand - the
collapse of $v_2$-flow of protons. Hence, the collapse is only possible
for a first order transition \protect\cite{Paech03}.}
\label{paech_ne_flow}
\end{figure}

A dramatic collapse of the flow $v_1$ is also observed by NA49
\cite{NA49_v2pr40}, again around 40 A$\cdot$GeV, where the collapse of
$v_2$ has been observed.  This is the highest energy - according to
\cite{Fodor04,Karsch04} and Fig. \ref{phasedia} - at which a first
order phase transition can be reached at the central rapidities of
relativistic heavy-ion collisions.  We, therefore, conclude that a
first order phase transition at the highest baryon densities accessible
in nature has been seen at these energies in Pb+Pb collisions.
Moreover, Fig. \ref{paech_ne_flow} shows that the elliptic flow clearly
distinguishes between a first order phase transition and a crossover.

\section{Strong collective flow at RHIC signals a new phase of matter}

\begin{figure}[h]
\psfig{file=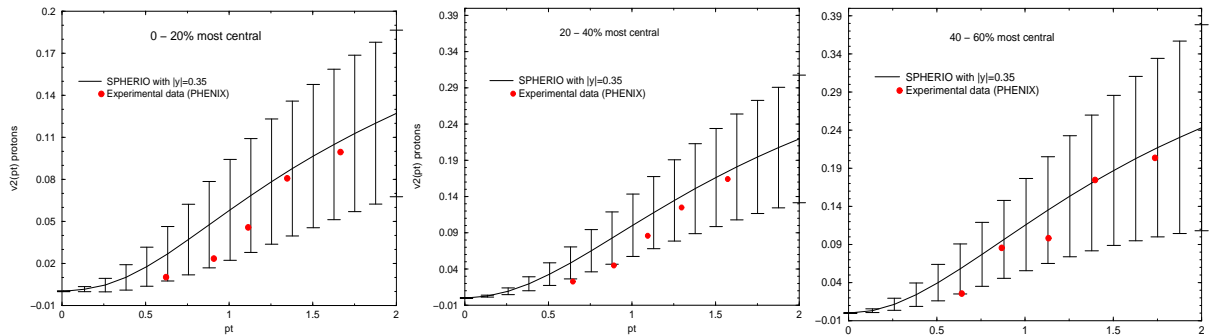,width=16cm}
\caption{\label{bernardo}Elliptic flow $v_2$ of protons as a function
of $p_T$ for different centralities.  NexSpherio ideal hydro results
exhibit about a factor of two higher proton-$v_2$-values than the
PHENIX data up to $p_T=1$ GeV/c. Furthermore, the fluctuations in the
NEXUS initial conditions for the SPHERIO- ideal hydro simulation
reflect  $>50$~\% event-by-event fluctuations of the proton-$v_2$
values.  Only large viscosity effects can damp out the too large $v_2$
flow itself as well as the fluctuations in $v_2$.}
\end{figure}

The rapid three-body thermalization found by Xu and Greiner (cf.
Sec. IV.A) justifies {\it a posteriori} the use of
hydrodynamical calculations for the time evolution of the complex
four-dimensional expansion of the plasma. However, there is no
justification for the use of simple ideal hydrodynamics (i.e.
neglecting the important transport coefficients) and simple, smooth
initial conditions in hydrodynamics \cite{Muronga01,Muronga03,Teaney}.
PHOBOS data at $\sqrt{s}$= 130 GeV and 200 GeV suggest energy
independent $v_2 (\eta)$ distributions.  Furthermore, the observed
distribution has a triangular shape.  This finding is in strong
disagreement with Bjorken boost invariant hydro predictions
\cite{Heinz04,Shuryak}, which fit only the midrapidity region.  The
predicted average proton $v_2$-values obtained from the SPHERIO hydro
code  with NEXUS initial conditons (Fig. \ref{bernardo},
\cite{Aguiar01}) are by factors of two higher than simple smooth
initial state hydrodynamic calculations.  This indicates that ideal
hydro with naive smooth initial conditions -- as used by many authors
-- do not describe but rather fit the data.  Strong viscosity effects
can play a role for particles with $p_T < 1.2$ GeV/c: a decent
description of the dynamics requires, however, relativistic viscous
hydro simulations \cite{Muronga01,Muronga03,Muronga03b}.  The
NexSpherio simulations (Fig. \ref{bernardo}, \cite{Aguiar01}) predict
very large event-by-event fluctuations of $v_2$ caused by the strongly
fluctuating initial conditions (given by NEXUS).  Is this in accord
with data?  What about the effect on the event plane determination?

\begin{figure}[!]
\centerline{\epsfig{file=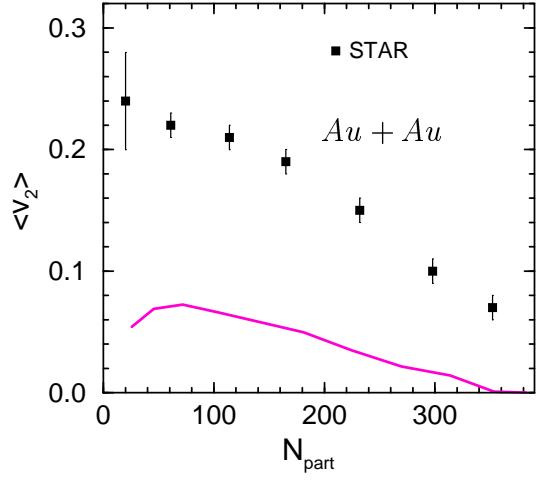,scale=0.85}}
\caption{High $p_T$- $v_2$ values as a function of $N_{part}$ as
measured by STAR are compared to HSD calculations. The $v_2$ data are
more than 5 times higher than the HSD model predictions for the most
central collisions \protect{\cite{CGG}}.}
\label{v2_rhic}
\end{figure}

\noindent
Microscopic transport simulations of particle yields, $dN/dy$
distributions, etc.  give a good description of the RHIC Au+Au data
\cite{Brat03}. The HSD and UrQMD transport approaches are based on
string, quark, diquark ($q, \bar{q}, qq, \bar{q}\bar{q}$) as well as
hadronic degrees of freedom. At RHIC, UrQMD and HSD yield reasonable
abundances of light hadrons composed of $u,d,s$ quarks
\footnote{For a more recent survey on hadron rapidity distributions
from 2 to 160 A$\cdot$GeV in central nucleus-nucleus collisions within the
HSD and UrQMD transport approaches we refer the reader to Ref.
\cite{Weber02}.}.
Do they also predict the collective flow properly?
The UrQMD prediction is clearly not compatible with the measured ~6\%
elliptic flow - it is sizeably underestimated \cite{Bleicher00}.  When
shortening the formation time \cite{Bleicher00} one can  get the model
results closer to the data (Fig. \ref{v2_rhic}) but more additional
initial pressure -- needed to create the missing extra flow -- is not
justified in  the hadronic transport models. At high transverse momenta
($p_T \approx 2$ GeV/c) the $v_2$-flow is underestimated even by a factor
of three (Fig. \ref{v2_rhic}) in the HSD model \cite{CGG}.  We mention
that the microscopic quark-gluon-string model inserts in addition short
distance vector repulsion in order to achieve high flow values
\cite{Zabrodin04}.

\begin{figure}[!]
\centerline{\psfig{file=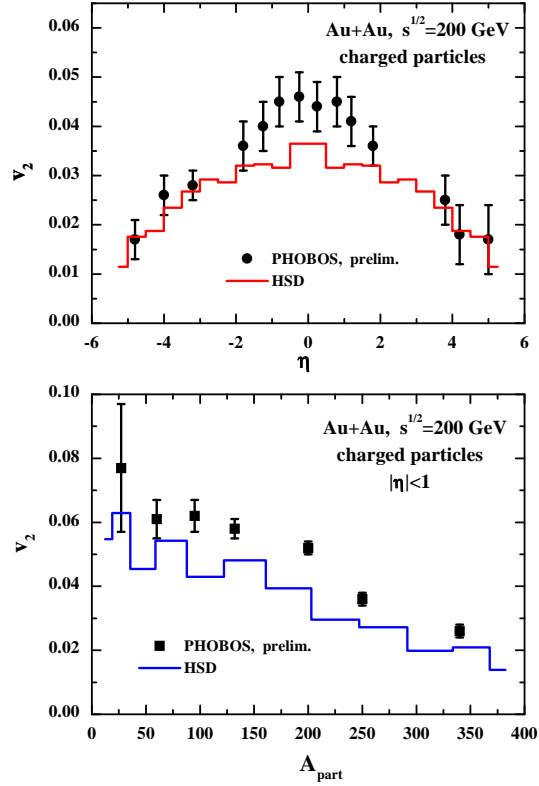,width=7cm}}
\caption{The elliptic flow $v_2$ for charged hadrons (HSD, solid
lines) as a function of pseudorapidity $\eta$ (upper part) and as a
function of the number of 'participating nucleons' $A_{part}$ for
$|\eta| \leq$ 1 (lower part) for $Au+Au$ collisions at $\sqrt{s}$ = 200
GeV, in comparison to the preliminary 'hit-based analysis' data of the
PHOBOS Collaboration \protect\cite{PHOBOS1}. Note, that in spite of the
shortened formation, $\tau \approx 0.8$ fm/c, HSD clearly
underpredicts the data even at the moderate $p_T$-values dominating
the $p_T$ integrated $v_2$-values shown here. At higher $p_T$-values,
Fig. \ref{v2_rhic}, the discrepancy to the data is more dramatic.}
\label{bild4}
\end{figure}

Is a transport approach -- based on strings and hadronic degrees of
freedom -- adequate in the initial stage of nucleus-nucleus collisions
at RHIC energies, where the quark-gluon plasma is formed? Well, the
particle abundancies show a rather smooth evolution  from SIS to RHIC
energies \cite{survey,Baker}. However, the effective partonic degrees
of freedom in the initial phase are needed to supply the large pressure
obviously needed to describe the elliptic flow at RHIC energies.  Even
'early' hadron formation -- as in HSD with $\tau = 0.8$ fm/c --
and 'large' (pre-)hadronic interaction cross sections are insufficient
to explain the $v_2$ flow data.  This is demonstrated in Fig.
\ref{bild4}, which shows the calculated elliptic flow $v_2$ from HSD
for charged hadrons (solid lines) as a function of the pseudorapidity
$\eta$ (upper part) and as a function of the number of 'participating
nucleons' $N_{part}$ (lower part)  for $|\eta| \leq$ 1 in comparison to
the preliminary 'hit-based analysis' data of the PHOBOS Collaboration
\cite{PHOBOS1}. The HSD results are very similar to those of the
hadronic rescattering model by Humanic et al.  \cite{Tom1,Tom2}
and agree with the calculations by Sahu et al.  \cite{Sahu02} performed
within the hadron-string cascade model JAM \cite{JAM}.

\section{Early Thermalization at RHIC - evidence for a new phase}

\subsection{Elastic and inelastic multi-particle collisions in a parton
cascade}

\begin{figure}[!b]
\vspace*{-5mm}
\centerline{\epsfig{file=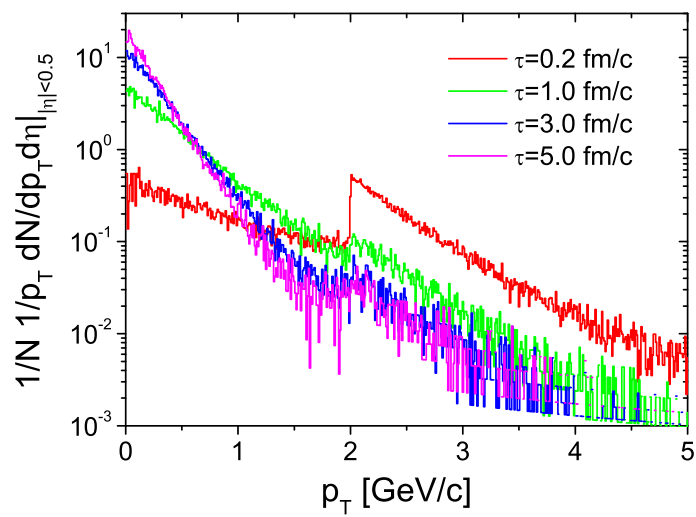,scale=0.95}}
\vspace{-0.5cm}
\caption{\label{minijet} Time evolution of the midrapidity transverse
momentum spectrum for a central RHIC collision from the
three-dimensional Xu-Greiner-parton cascade simulation
\cite{Xu04,XuDiss} (prelimenary results are from \cite{Greiner04} -
including transverse expansion).  Only partons residing in a central
cylinder of radius $R\leq 5$ fm are plotted.  The initial
off-equilibrium conditions are given by a minijet distribution with
corresponding overlap function in space-time.  At $t=0$, only minijets
with $p_T>2$ GeV/c are populated.  Energy degradation to lower momenta
proceeds by rapid gluon emission within the first fm/c.
(Quasi-)kinetic and chemical equilibrium is found up to 4 fm/c.  Here
longitudinal and transversal hydrodynamic work is at action, resulting
in a rapid lowering of the temperature by a factor of two.  On the
other hand, a small fraction of the initial non-equilibrium high
momentum power-law tail of the mini-jet production survives even in
this central cell.  } \end{figure}

To describe the early dynamics of ultrarelativistic heavy-ion
collisions and to address the crucial question of thermalization and
the early pressure build up at RHIC, unexpectedly high elastic parton
cross sections  have been assumed in parton cascades \cite{Texas,MG02}
in order to reproduce the elliptic flow $v_2(p_T)$ seen experimentally
at RHIC.  These cross sections are about 1/9 of the baryon-baryon total
cross section ($\sim$ 45 mb) or 1/6 of the meson-baryon cross section
($\sim$ 30 mb), such that the effective cross section for the
constituent quarks and antiquarks is roughly the same in the partonic
and hadronic phase, however, tenfold higher than the cross sections
calculated in pQCD.

It has been a great puzzle until recently, when Xu and Greiner
developed a consistent three-body kinetic parton cascade algorithm
\cite{Xu04,XuDiss}.  These stochastic inelastic (`Bremsstrahlung') $2
\rightarrow 3$ and $3 \rightarrow 2$ collisions ($gg \leftrightarrow
ggg $) drive early thermalization, rather than the two-body elastic
collisions, which are too strongly forward peaked.  A quantitative
understanding of the early dynamical stages of deconfined matter is
finally in sight.  Parton cascade analyses incorporating only binary $2
\leftrightarrow 2$ pQCD scattering processes can not build up
thermalization and early quasi-hydrodynamic behaviour necessary for
achieving sufficient elliptic flow.  The importance of inelastic
reactions was raised in the so called `bottom-up thermalization'
picture \cite{B01}; gluon multiplication leads to a much faster
equilibration.

Xu and Greiner consider - besides gluon- and quark- two-body
elementary parton-parton scatterings - three-body processes $gg
\leftrightarrow ggg $ in leading-order pQCD. They employ effective
Landau-Pomeranchuk-Migdal suppression and standard screening masses for
the infrared sector of the scattering amplitude.

The early stage of gluon production in the Xu-Greiner (X-G) approach
(Fig. \ref{minijet}) leads to a rapid kinetic equilibration of the
momentum distribution as well as to a rather abrupt lowering of the
temperature by soft gluon emission. Detailed balance among gain and
loss contributions is reached rapidly, too.  The later, slower time
evolution is then governed by chemical equilibration of the quark
degrees of freedom.  The X-G cascade does allow to study in detail
RHIC- collisions with various initial conditions like minijets or color
glass condensate (CGC). Fig.~\ref{minijet} depicts a preliminary
calculation \cite{Greiner04} using minijet-initial conditions.

Thermalization and chemical equilibration -- as proposed in the
bottom-up scenario \cite{B01} -- can thoroughly be tested within this
approach.  The impact parameter dependence on the transverse energy is
used to understand elliptic and transverse flow at RHIC within this new
kinetic parton cascade with inelastic $3\rightarrow 2$ and
$2\rightarrow 3$ interactions.

\section{How much quenching of high $p_T$ hadrons is due to (pre-) hadronic
final state interactions?}

A (mini-)jet at RHIC can produce hard particles, with $p_T$ above 5
GeV/c, but must also form soft particles with $p_T$ around 2 GeV/c.  Jets
produced in the center of the plasma zone have to pass first through
the parton phase at very high temperatures, then through the correlated
diquark and constituent quarks and finally through the  hadronic phase
that has build up preferentially close to the surface of the fireball.
Very high $p_T$ jets with $\gamma > 10$ materialize only far outside of
the plasma.  Most of the jets -- observed at RHIC -- are at $p_T
\approx 4-5$~GeV/c.  More than 50\% of the leading jet particles at $p_T
\sim 5\,$GeV/c are baryons.  Pion jets of 5 GeV have a $\gamma = 35$,
i.e., they form far outside the plasma.  However,
HSD-PYTHIA-calculations \cite{Cassing04} show that many pions stem from
decaying rho-jets.  But, $\rho$'s and protons of 5 GeV have $\gamma =
5$. Thus, $\rho$ and p-jets hadronize with roughly 50\% probability
\cite{Kai,CGG} while passing through the expanding bulk matter.  All
partonic and hadronic models have failed by factors of 5-10 to predict
the observed high baryon abundance.

The PHENIX \cite{PHENIX1} and STAR \cite{STAR1} collaborations reported
a suppression of meson spectra for transverse momenta $p_T$ above $\sim
3\,$GeV/c.  This suppression is not observed in d+Au interactions at the
same bombarding energy per nucleon \cite{E1,E2} and presents clear
evidence for the presence of a new form of matter.  However, it is not
clear at present how much of the observed suppression can be attributed
to (pre-)hadronic interactions (FSI) \cite{Kai}.  (In-)elastic
collisions of (pre-)hadronic high momentum states with some of the bulk
(pre-)hadrons in the fireball can contribute in particular to the
attenuation of $p_T \approx 5\,$GeV/c  transverse momentum hadrons at
RHIC \cite{Cassing04}:  Most of the medium momentum (pre-)hadrons from
a $\pm 5$ GeV/c double jet will materialize inside the dense plasma;
their transverse momenta being 0-4 GeV/c.  The particles are dominantly
$\rho$'s, K's and baryons at $p_T >2.5\,$GeV/c -- hence their formation
time is $\gamma \tau_F \approx 4$ fm/c in the plasma rest frame.  The
time for color neutralization can also be very small \cite{Kopel4} for
the leading particle due to early gluon emission.

The (pre-)hadronic interactions with the bulk of the (pre-)hadronic
comovers then must have clearly an effect: they, too, suppress the
$p_T$-spectrum \cite{Kai}.  (In)elastic reactions of the fragmented
(pre-)hadrons with (pre-)hadrons of the bulk system cannot be described
by pQCD:  The relevant energy scale $\sqrt{s}$ is a few GeV.  Such
(in-)elastic collisions are very efficient for energy degradation since
many hadrons with lower energies are produced.  On the average, 1 to 2
such interactions can account for up to 50\% of the attenuation of high
$p_T$ hadrons at RHIC \cite{Kai}.  Hence, the  hadronic fraction of the
jet-attenuation  had to be addressed.

In Ref. \cite{CGG} the HSD transport approach \cite{Cassing99} is
employed. Moderate to high transverse momenta ($>1.5$ GeV/c) are
incorporated by a superposition of $p+p$ collisions described via
PYTHIA \cite{PYTHIA}.  In Au+Au collisions, the formation of secondary
hadrons is not only controlled by the formation time $\tau_f$, but also
by the energy density in the local rest frame. In \cite{CGG}, hadrons
are not allowed to be formed if the energy density is above
1 GeV/fm$^3$\footnote{This energy density cut is employed in the
default HSD approach.}.  The interaction of the leading and energetic
(pre-)hadrons with the soft hadronic and bulk matter is thus explicitly
modeled.

\begin{figure}[t]
\centerline{\epsfig{file=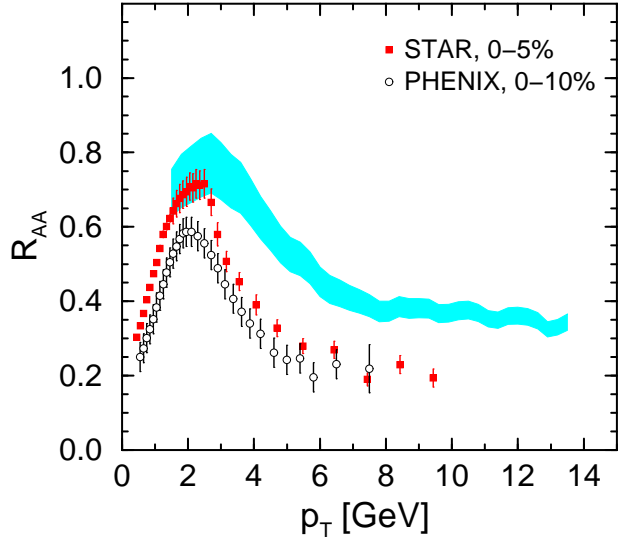,scale=0.45,angle=-90}}
\caption{The suppression
factor $R_{AA}$ (3) of charged hadrons
      at 5\% (10\%) central $Au+Au$ collisions ($\sqrt{s}$=200 GeV)
      at midrapidity  (hatched band).
      The experimental data are from Refs. \cite{PHENIX2,STAR} and show
      clearly that an additional  partonic suppression is needed
      (taken from Ref. \protect\cite{CGG}).
} \label{fig5}
\end{figure}

\begin{figure}[b]
\centerline{\epsfig{file=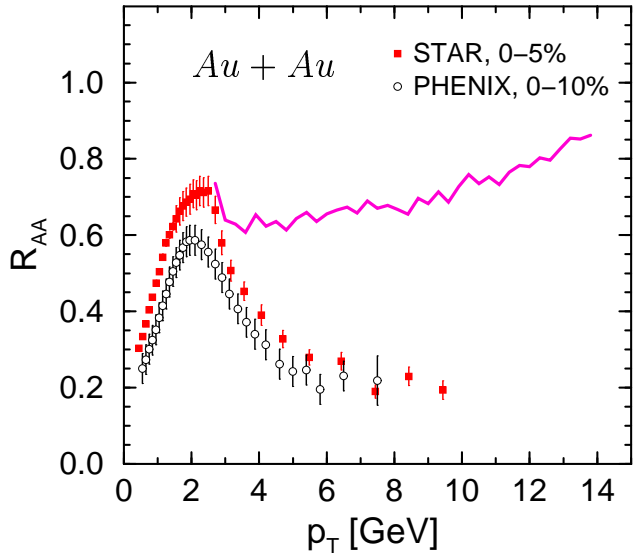,scale=1}}
    \caption{
      Same as Fig.~\ref{fig5}, but with a leading cross section
      according to eq.~(\ref{sigma}) for the perturbative high $p_T$
      particles. The observed attenuation is more than double the value!
      }
    \label{fig10}
\end{figure}

Figs.~18, 19 show the nuclear modification
factor \cite{Cassing04}
\begin{equation}
  \label{ratioAA}
  R_{\rm AA}(p_T) = \frac{1/N_{\rm AA}^{\rm event}\ d^2N_{\rm AA}/dy dp_T}
  {\left<N_{\rm coll}\right>/\sigma_{pp}^{\rm inelas}\ d^2
    \sigma_{pp}/dy dp_T}\ .
\end{equation}
for the most central (5\% centrality)
Au+Au collisions at RHIC.  The Cronin enhancement is visible at all
momenta.  Hadron formation time effects do play a substantial role in
the few GeV region, since heavier hadrons (K$^*$'s, $\rho$'s, protons)
are formed 7 times earlier than the rather light pions in the cms frame
at fixed transverse momentum due to the lower Lorentz boost $\gamma <5$.
It was shown in \cite{CGG} that for transverse momenta $p_T \geq 6$ GeV/c
the interactions of formed hadrons are not able to explain the
attenuation observed experimentally. However, the ratio $R_{\rm AA}$ is
influenced by interactions of formed (pre-)hadrons in the $p_T =
1\dots5\,{\rm GeV}/c$ range \cite{CGG}.  A similar behaviour has also
been found in UrQMD simulations \cite{Bass_UrQMD}.

As pointed out before, the suppression seen in the calculation for
larger transverse momentum hadrons is due to the interactions of the
leading (pre-)hadrons with target/projectile nucleons and the bulk of
low momentum hadrons. It is clear that the experimentally observed
suppression can not be quantitatively described by the (pre-)hadronic
attenuation of the leading particles \cite{CGG}.  The ratio $R_{\rm
AA}$ (3) decreases to a value of about $0.5$ at 5 GeV for central
collisions, whereas the data are around $R_{\rm AA}\approx 0.25$.

To check, how robust this HSD estimate is,
alternative models for the leading pre-hadron cross
section have been studied in \cite{CGG} by
adopting a time--dependent, color-transparency-motivated
cross section for
leading pre-hadrons \cite{Gerland}
\begin{equation}
  \label{sigma}
  \sigma_{\rm lead}(\sqrt{s},\tau) = \frac{\tau-\tau_0}{\tau_f} \sigma_{\rm had}
  (\sqrt{s})
\end{equation}
for $\tau-\tau_0 \leq \tau_f$, where $\tau_0$ denotes the actual
production time, $\tau_f$ the formation time, after which the full
hadronic cross section is reached.

Within this scenario the attenuation is 35\% at $p_T\sim5\,{\rm GeV}/c$
(see Fig. \ref{fig10}), while the data show more than double the
attenuation.  Thus, (pre-)hadronic jet interactions cannot provide a
quantitative explanation for the jet suppression observed.  They do
provide, however, a sizable ($30-50 \%$) contribution to the jet
quenching.

For particles observable with momenta $p_T \geq 4$ GeV/c, the HSD
transport calculation predicts that still 1/3 of the final observed
hadrons have suffered one or more interactions, whereas the other 2/3
escape freely, i.e., without any interaction (even for central
collisions).  This implies that the final high $p_T$ hadrons originate
basically from the surface.

\subsection{Angular Correlations of Jets -- Can jets fake the large
$v_2$-values observed?}

\begin{figure}[h!]
\phantom{a}\vspace*{-5mm}
    \includegraphics[angle=-90,width=8cm]{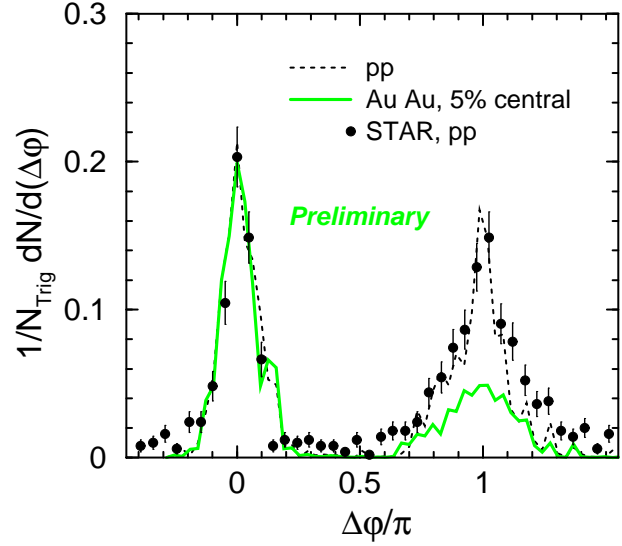}
    \caption{
\vspace*{2mm}
     STAR data on near-side and away-side jet correlations compared
     to the HSD model for p+p and central
Au+Au collisions at midrapidity for $p_TN{Trig}=4\dots6\,{\rm GeV}/c$ and
$p_T=2\,{\rm GeV}/c\dots p_TN{Trig}$. \protect{\cite{CGG,Cassing04}}
      }
    \label{angcorr}
\end{figure}

Fig.~\ref{angcorr} \cite{Cassing04} shows the angular correlation of
high $p_T$ particles ($p_T N{Trig}=4\dots6\,{\rm GeV}/c$,
$p_T=2\,{\rm GeV}\dots p_T N{Trig}$, $|y| <0.7$) for the 5\% most
central Au+Au collisions at $\sqrt{s}$ = 200 GeV (solid line) as well
as $pp$ reactions (dashed line) from the HSD-model \cite {Cassing04} in
comparison to the data from STAR for $pp$ collisions
\cite{StarAngCorr}.  Gating on high $p_T$ hadrons (in the vacuum)
yields 'near--side' correlations in Au+Au collisions close to the
'near--side' correlations observed for jet fragmentation in the vacuum
(pp).  This is in agreement with the experimental observation
\cite{StarAngCorr}.  However, for the away-side jet correlations,  the
authors of Ref. \cite {Cassing04} get only a  $\sim$50\% reduction,
similar to HIJING, which has only parton quenching and neglects hadron
rescattering.  Clearly, the observed \cite{StarAngCorr} complete
disappearance of the away-side jet (Fig.~\ref{angcorr}) cannot be
explained in the HSD-(pre-)hadronic cascade even with a small formation
time of $0.8\,$fm/c. Hence, the correlation data provide another  clear
proof for the existence of the bulk plasma.

Although (pre-)hadronic final state interactions yield a sizable ($\leq
50 \%$) contribution to the high $p_T$ suppression effects observed in
Au+Au collisions at RHIC, $\sim 50 \%$ of the jet suppression
originates from interactions in the plasma phase.  The elliptic flow,
$v_2$, for high transverse momentum particles is underestimated by at
least a factor of 3 in the HSD transport calculations \cite{CGG} (cf.
Fig.~\ref{v2_rhic}).  The experimentally observed proton excess over
pions at transverse momenta $p_T > 2.5$ GeV/c cannot be explained
within the CGG approach \cite{CGG}; in fact, the proton yield at high
$p_T \geq 5$ GeV/c is a factor 5-10 too small.  We point out that this
also holds for partonic jet-quenching models.  Further experimental
data on the suppression of high momentum hadrons from d+Au and Au+Au
collisions, down to $\sqrt{s}$ = 20 GeV, are desperately needed to
separate initial state Cronin effects from final state attenuation and
to disentangle the role of partons in the colored parton plasma from
those of interacting pre-hadrons in the hot and dense fireball.

Can the attenuation of jets of $p_T \ge5\,$GeV/c actually fake the
observed $v_2$-values at $p_T \approx 2\,$GeV/c? This question comes
about since due to fragmentation and rescattering a lot of
momentum-degraded hadrons will propagate in the hemisphere defined by
the jets. However, their momentum dispersion perpendicular to the jet
direction is  so large that it could indeed fake a collective flow that
is interpreted as coming from the high pressure early plasma phase.

\begin{figure}[h!]
    \includegraphics[width=9cm]{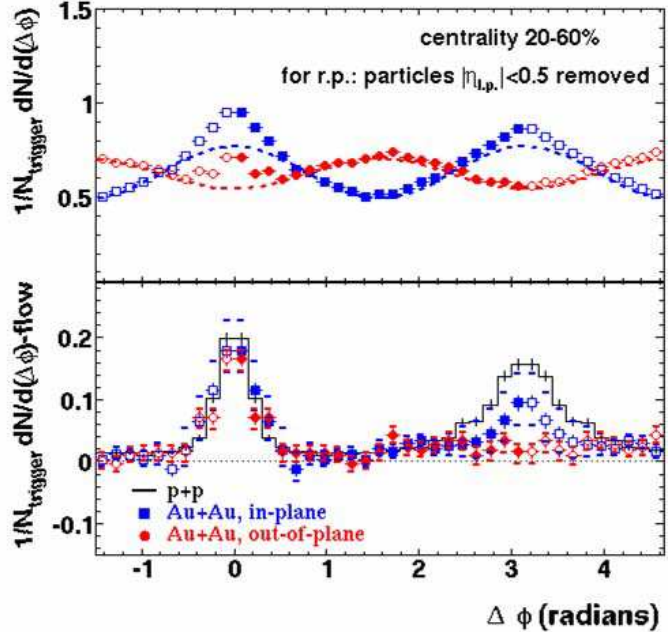}
\caption{High $p_T$ correlations: in-plane vs. out-of-plane
correlations of the probe (jet+secondary jet fragments) with the bulk
($v_2$ of the plasma at $p_T > 2\,$GeV/c), prove the existence of the
initial plasma state (STAR-collaboration, preliminary).}
\label{filimonov}
\end{figure}

\begin{figure}[h!]
\begin{center}
\includegraphics[width=8cm]{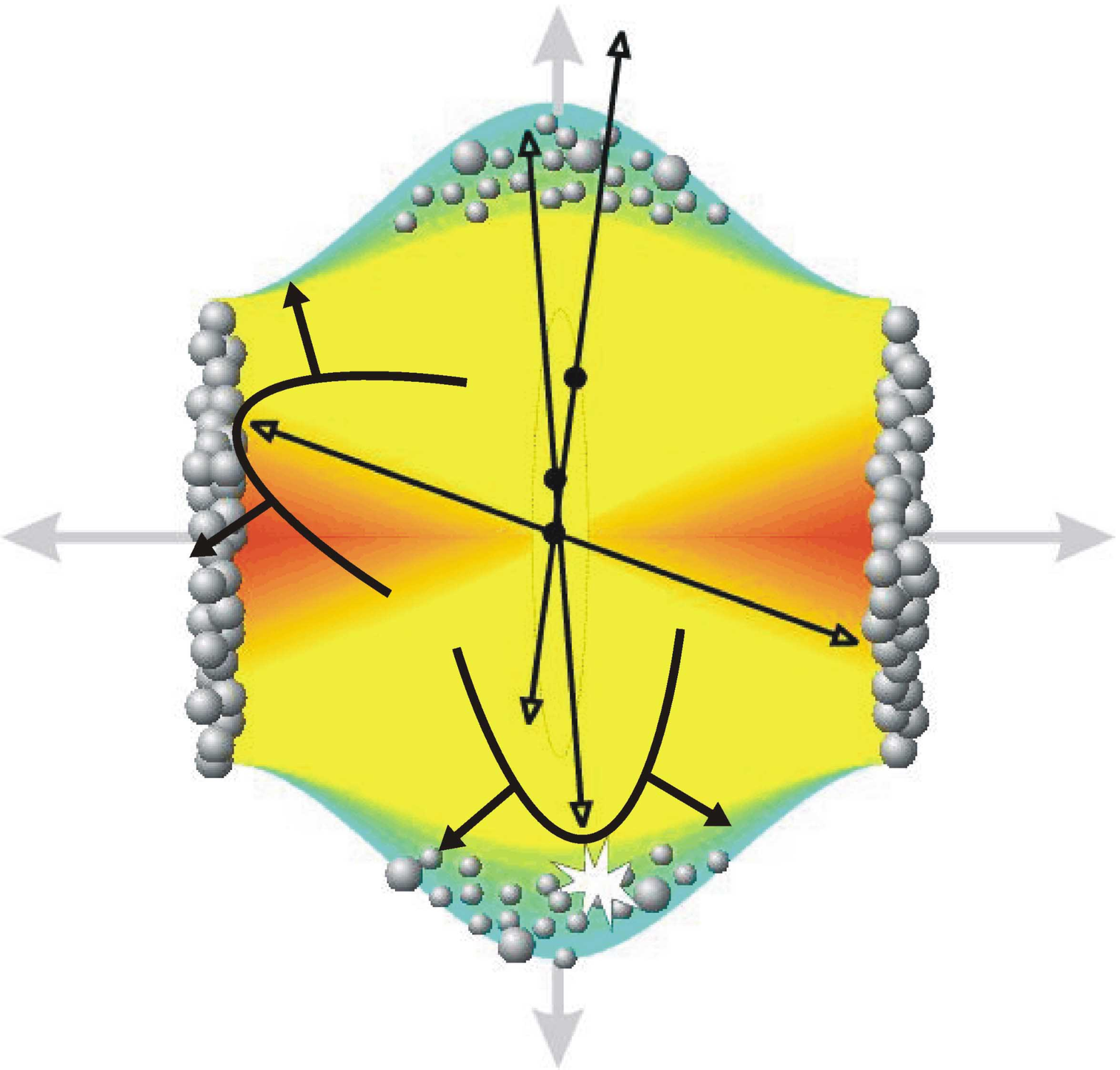}
\end{center}
\caption{Illustration of jets traveling through the late hadronic stage
of the reaction.  Only jets  from the region close to the initial
surface can propagate and fragment in the vacuum
\protect{\cite{Hofmann74,Cassing04,Baum75,Rischke90}}.  The other jets
will interact with the bulk, resulting in wakes with bow waves
travelling transversely to the jet axis.}
\label{fig:scheme}
\end{figure}

On first sight, Fig. \ref{filimonov} shows that this could indeed  be
the case:  the in-plane $v_2$ correlations are aligned with the jet
axis, the away-side bump, usually  attributed to collective $v_2$ flow
(dashed line), could well be rather due to the stopped, fragmented and
rescattered away-side jet!  However, this argument is falsified by the
out-of-plane correlations (circles in Fig. \ref{filimonov}).  The
near-side jet is clearly visible in the valley of the collective flow
$v_2$ distribution.  Note that $v_2$ peaks at $\varphi = \pi/2$
relative to the jet axis! The away-side jet, on the other hand, has
completely vanished in the out-of-plane distribution
(cf. Fig. \ref{fig:scheme})!

Where are all the jet fragments gone?  Why is there no trace left?
Even if the away-side jet fragments completely and the fragments get
stuck in the plasma, leftovers should be detected at momenta below
$2\,$GeV/c.  Hadronic models as well as parton cascades will have a
hard time to get a quantitative agreement with these exciting data!


We propose future correlation measurements which can yield
spectroscopic information on the plasma.

\begin{enumerate}
\item
If the plasma is a colorelectric plasma, experiments will - in spite of
strong plasma damping - be able to search for wake-riding potential
effects. The wake of the leading jet particle can trap comoving
companions that move through the plasma in the wake pocket with the
same speed ($p_T/m$) as the leading particle. This can be particular
stable for charmed jets due to the deadcone effect as proposed by
Kharzeev et al \cite{Kharzeev}, which will guarantee little energy
loss, i.e. constant velocity of the leading D-meson. The leading
D-meson will practically have very little momentum degradation in the
plasma and therefore the wake potential following the D will be able to
capture the equal speed companion, which can be detected
\cite{Schafer78}.

\item
One may measure the sound velocity of the expanding plasma by the
emission pattern of the plasma particles travelling sideways with
respect to the jet axis: The dispersive wave generated by the wake of
the jet in the plasma yields preferential emission to an angle
(relative to the jet axis) which is given by the ratio of the leading
jet particles' velocity, devided by the sound velocity in the hot dense
plasma rest frame.  The speed of sound for a non-interacting gas of
relativistic massless plasma particles is $c_s \approx
\frac{1}{\sqrt{3}} \approx 57 \% \,c$, while for a plasma with strong
vector interactions, $c_s =  c$.  Hence, the emission angle measurement
can yield information of the interactions in the plasma.
\end{enumerate}

\section{Summary}

The NA49 collaboration has observed the collapse of both, $v_1$- and
$v_2$-collective flow of protons, in Pb+Pb collisions at 40 A$\cdot$GeV,
which presents first evidence for a first order phase transition in
baryon-rich dense matter. It will be possible to study the nature of
this transition and the properties of the expected chirally restored
and deconfined phase both at the forward fragmentation region at RHIC,
with upgraded and/or second generation detectors, and at the new GSI
facility FAIR.  According to Lattice QCD results
\cite{Fodor04,Karsch04}, the first order phase transition occurs for
chemical potentials above 400 GeV.  Fig. \ref{paech_ne_flow} shows that
the elliptic flow clearly distinguishes between a first order phase
transition and a crossover.  Thus, the observed collapse of flow, as
predicted in \cite{Hofmann74,Hofmann76}, is a clear signal for a first
order phase transition at the highest baryon densities.

A critical discussion of the use of collective flow as a barometer for
the equation of state (EoS) of hot dense matter at RHIC showed that
hadronic rescattering models can explain $< 30 \%$ of the observed
flow, $v_2$, for $p_T > 2$ GeV/c.  We interpret this as evidence for
the production of superdense matter at RHIC with initial pressure way
above hadronic pressure, $p > 1$~GeV/fm$^3$.

The fluctuations in the flow, $v_1$ and $v_2$, should be measured.
Ideal Hydrodynamics predicts that they are larger than 50 \%  due to
initial state fluctuations.  The QGP coefficient of viscosity may be
determined experimentally from the fluctuations observed.

The connection of $v_2$ to jet suppression is examined. It is proven
experimentally that the collective flow is not faked by minijet
fragmentation and that the away-side jet suppression can only partially
($<$ 50\%) be due to pre-hadronic or hadronic rescattering.

We propose upgrades and second generation experiments at RHIC, which
inspect the first order phase transition in the fragmentation region,
i.e. at $\mu_B\approx~400$~MeV  ($y \approx 4-5$), where the collapse
of the proton flow analogous to the 40 A$\cdot$GeV data should be seen.

The study of Jet-Wake-riding potentials and Bow shocks caused by jets
in the QGP formed at RHIC can give further clues on the equation of
state  and transport coefficients of the Quark Gluon Plasma.

\vspace*{3mm}
\begin{acknowledgements}
I like to thank
E. L. Bratkovskaya, M. Bleicher, A. Muronga, K. Paech, M. Reiter,
S. Scherer, S. Soff, H. Weber, G. Zeeb, D. Zschiesche,
W. Cassing, C. Greiner, K. Gallmeister, Z. Xu,
B. Tavares, L. Portugal, C. Aguiar, T. Kodama, F. Grassi, Y. Hama,
T. Osada, O. Sokolowski, and K. Werner for their contributions that
have made this review possible.
\end{acknowledgements}


\end{document}